\documentclass[12pt, a4paper]{article}
\usepackage{spconf,amsmath,graphicx}
\usepackage{graphicx}
\usepackage{amssymb,amsmath,bm}
\usepackage{textcomp}
\usepackage{graphicx}
\usepackage{wrapfig}
\usepackage{lipsum}
\usepackage{dblfloatfix}
\usepackage{fixltx2e}
\usepackage{nameref}
\usepackage{amsmath}
\usepackage{hyperref}
\usepackage{graphicx}
\usepackage{varioref}
\usepackage{booktabs}  
\usepackage{siunitx}
\usepackage{numprint}
\usepackage{color}
\usepackage{textcomp}
\usepackage{geometry}
\usepackage{booktabs}  
\usepackage{lipsum}

\title{A Robust Diarization System for Measuring Dominance in Peer-Led Team Learning Groups}
\makeatletter
\def\name#1{\gdef\@name{#1\\}}
\makeatother \name{{\em Harishchandra Dubey\thanks{\textcolor{blue}{This material is presented to ensure timely dissemination of scholarly and technical work. Copyright and all rights therein are retained by the authors or by the respective copyright holders. The original citation of this paper is:
				H. Dubey, A. Sangwan, J. H.L. Hansen, A Robust Diarization System For
				Measuring Dominance in Peer-Led Team Learning Groups, IEEE Workshop on Spoken Language Technology 2016, December, 2016, San Diego, California, USA.}}, Abhijeet Sangwan, John H. L. Hansen\textsuperscript{+}\thanks{\textsuperscript{+}This project was funded in part by AFRL under contract FA8750-15-1-0205 and partially by the University of Texas at Dallas from the Distinguished University Chair in Telecommunications Engineering held by J. H. L. Hansen.}}}
\address{Center for Robust Speech Systems, Eric Jonsson School of Engineering\\
The University of Texas at Dallas, Richardson, TX 75080, USA \\
{\small \tt \{harishchandra.dubey, abhijeet.sangwan, john.hansen\}@utdallas.edu}
}
\begin{document}
%
\maketitle
\begin{abstract}
Peer-Led Team Learning (PLTL) is a structured learning model where a team leader is appointed to facilitate collaborative problem solving among students for Science, Technology, Engineering and Mathematics (STEM) courses. This paper presents an informed HMM-based speaker diarization system. The minimum duration of short conversational-turns and number of participating students were fed as side information to the HMM system. A modified form of Bayesian Information Criterion (BIC) was used for iterative merging and re-segmentation. Finally, we used the diarization output to compute a novel dominance score based on unsupervised acoustic analysis. 
\end{abstract}
\vspace{-2mm}
\section{Introduction}
\label{sec:intro}
\vspace{-2mm}
%
%
%
%
Peer-Led Team Learning (PLTL) is a strategy where student groups collaboratively solve problems for a given course. Such a session is usually coordinated by a peer leader, who has taken and passed the course in earlier semesters. PLTL has been adopted for various undergraduate Science, Technology, Engineering and Mathematics (STEM) courses, where it has shown positives outcomes towards learning~\cite{snyder2016peer}. The traditional teaching model lacks one-to-one interaction and peer-feedback unlike PLTL. Peer leaders are also expected to give helpful hints and comments during students' discussion. Peer leaders are not supposed to reveal solutions, in contrast to the traditional teaching model~\cite{cracolice2001peer}.

Analysis of PLTL team behavior using spoken language technology could also identify best practices in terms of team composition, early intervention, impact of various team parameters on outcome,~\emph{etc.}. For example, PLTL recordings could help in identifying students who are experiencing difficulty in learning a subject early on in the process. In the most general sense, PLTL is a small group meeting periodically and working towards a focused goal. Hence, various aspects of the group, such as team behavior, cohesion, productivity, sentiment,~\emph{etc.} are interesting topics to study. 
%
\begin{figure*}[!t]
\centering
\includegraphics[width=460pt]{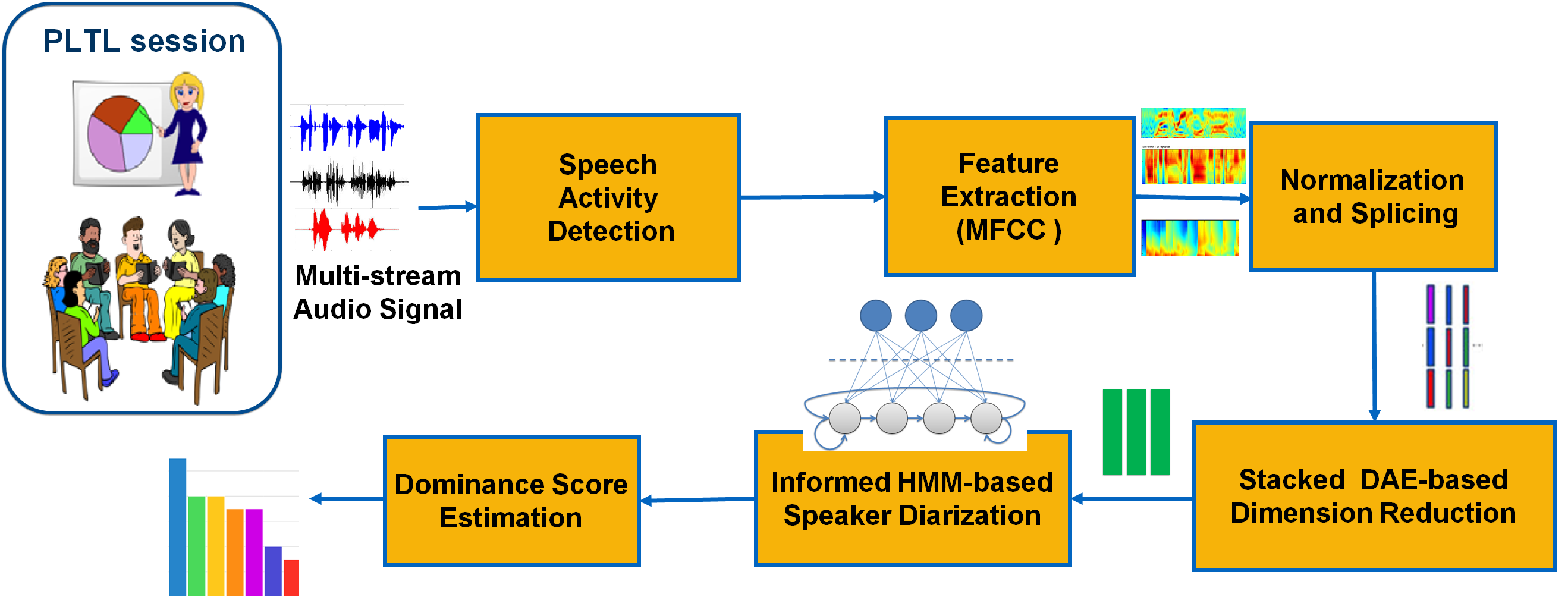}
\caption{The proposed system consisting of six stages: Speech Activity Detection (SAD), Feature Extraction (MFCC), Mean and Variance normalization of features and splicing of features using 5 frames of context from past and future, stacked Denoising Autoencoder (DAE)-based dimension reduction, Informed HMM-based diarization, and unsupervised estimation of dominance score.}
\label{fig_pltl}
\end{figure*}
%
%
%
%
Particularly, this study makes the following contributions in audio-based analysis of PLTL groups: 1) We propose a feature engineering technique that combines audio features from multiple audio streams. The new method uses stacked denoising autoencoders (DAE) for non-linear dimension reduction of spliced MFCC features from multiple audio-streams; 2) We also propose an informed HMM-based diarization system that accomplishes diarization via unsupervised joint segmentation and clustering; 3) A new method for estimating Dominance Score (DS) using unsupervised acoustic analysis; 4) A new technique for speaker energy computation using Wavelet Packet Energy (WPE); 5) The proposed methods were evaluated on PLTL sessions extracted from the CRSS-PLTL corpus~\cite{dubey2016interspeech}.
\vspace{-2mm}
\section{Proposed System}
\label{sec:proposed}
\vspace{-2mm}
In this section, we discuss the proposed diarization system that consists of stacked denoising autoencoders (DAE) for dimension reduction, and informed Hidden Markov Model (HMM) for joint segmentation and clustering. In the first step, we removed the non-speech (NS) frames from the audio signal, followed by extraction of MFCC features. The features are mean and variance normalized followed by time splicing using 5 context-frames from past and future. A stacked DAE system is then used to reduce the feature dimension using a bottleneck architecture (BNF)~\cite{gehring2013extracting}. Next, the HMM system uses BNF along with two dimensions of side information,~\emph{i.e.}, number of speakers and minimum duration of speaker-turns. Hence, we call the system as informed HMM system. The iterative diarization procedure has three steps: (i) initial segmentation,  (ii) merging, and (iii) re-estimation. It is discussed in detail later in Section~\ref{sec:hmm}. The CRSS-PLTL Corpus used for evaluation of proposed algorithms was introduced in our earlier work~\cite{dubey2016interspeech}.
\vspace{-2mm}
\subsection{Bottleneck Features: Denoising Autoencoder (DAE)-based Dimension Reduction}
\label{sec:dae}
\vspace{-2mm}
%
%
%
%
%
%
We used 13-dimensional Mel-Frequency Cepstral Coefficients (MFCC) for extracting features from each frame. The parameters of the proposed system are given in Table~\ref{table_params}. The MFCC features were first mean-and-variance normalized. Since all the channel were delayed and scaled versions of the same speech signal at a given frame, we concatenated the normalized MFCC features from each channel (7) to form a feature super-vector (91=7*13 dimensional). Next, we used splicing for context of 5 past and 5 future frames. Thus, the spliced feature dimension become 1001(=91*11). The splicing incorporates the long-term context leading to a better representation of multi-stream speech data.

Denoising Autoencoders (DAE) were found useful in the dimension reduction tasks~\cite{wang2016auto}. DAE is trained in a way that allows it to learn low-dimensional hidden representation of the data such that, taking noisy input, it could reconstruct the input. The spliced features were corrupted with additive random noise before feeding it into the stacked DAE and it was trained to minimize the reconstruction error with respect to original input. The high dimensional spliced features (1001) necessitated the dimension reduction by taking the bottleneck features (BNF) from a stacked DAE. The parameters of the stacked DAE used for dimension reduction are given in Table~\ref{table_params}. Several denoising autoencoders (DAE) were stacked to form a 5 layered deep network. We used the PDNN toolkit~\cite{miaopdnn} with corruption parameter 0.2 and learning rate, momentum factor parameters of 0.01 and 0.05 respectively for our system. 
\vspace{-2mm}
\subsection{Informed HMM-based Diarization}
\label{sec:hmm}
\vspace{-2mm}
%
%
%
The diarization for PLTL sessions is different with respect to information available such as speaker count and turn statistics. The rapid short-turns, overlapped speech, and huge amount of reverberation and noise make the task challenging. Most of the diarization system studied did not address such challenges~\cite{dubey2016interspeech}. HMMs had been used in previous studies for various audio segmentation tasks in varied forms~\cite{kotti2008speaker,ajmera2002improved,huang2006advances2}. However, using side information, application to PLTL session, and using stacked DAE-based BNF are novel contributions of this paper with respect to diarization. Initially, we performed over-segmentation by dividing speech into $OS$ segments where $OS$ is 3 to 6 times the expected number of speakers. A HMM with $OS$ states is assumed for initial segments. Each HMM state has an output probability density function (PDF) that was modeled by $M$ component Gaussian Mixture Model (GMM). Each state of HMM was allowed to have $T$ sub-states to incorporate the minimum duration constraint. All sub-states of a given HMM state (hypothesized speaker cluster) share the GMM corresponding to their state. The HMM system was trained using the~\textit{Expectation-Maximization} (EM) algorithm. Once HMM was trained, we obtained the Viterbi path for all frames. Next, we used the Viterbi path for checking the binary merging hypothesis based on modified $G^{3}$ algorithm~\cite{dubey2016interspeech}. After the merge iteration finished, a new HMM with fewer states was trained. The whole process was repeated again until it converged based on two conditions. The first condition is to stop merging once the number of HMM states is equal to the number of speakers, and second one is to get no improvements in likelihoods ratio upon merging. 

We performed merging based on $G^{3}$ algorithm that is a variant of BIC and eliminates the need of a threshold (penalty term).  This trick was first developed to improve the speaker change detection as compared to BIC~\cite{ajmera2004robust}. In this paper, we use the same modelling techniques for a slightly different binary hypothesis to decide merging of two over-segmented segments or equivalently two HMM states. There are some modifications to $G^{3}$ algorithm applied for merging most-similar segments (HHM states). First, the minimum duration of staying in a HMM state or segment is much lower, 0.5s to 1s owing to the rapid short conversational-turns. The initial segments were modeled with a Gaussian Mixture with only $M_{s}$ components. After merging two initial segments, each modeled with $M_{s}$ components, the merged segment is modeled with $2M_{s}$ components. In this way, the number of parameters of the GMM model for merged segment is same as the sum of number of parameters in child segments. As a result of keeping the number of parameters the same at each merging step, we eliminated the BIC penalty term. Once the merging is done, the new HMM of smaller size is estimated where the GMM for each state is re-estimated using the EM algorithm. The acoustic features belonging to that HMM state (speaker) were used to re-estimate the corresponding GMM.
\vspace{-2mm}
\subsection{Speaker Energy Using Wavelet Packet Decomposition}
\label{sec:wpd}
\vspace{-2mm}
%
%
Earlier we had used formant energy for computing the speaker energy~\cite{dubey2016interspeech}. Formant energy was robust to the noise and distortions as compared to energy computed using short-time spectrum, at the expense of huge computational requirement. In this paper, we employed wavelet packet decomposition (WPD) for estimating the speaker energy. The wavelet packets (WPs) provide good time-frequency resolution at reasonable computational load~\cite{wickerhauser1991lectures}. There are several computationally simple methods for estimating WPs. We added the squared WP coefficient corresponding to the frequency range [50, 2000] Hz for capturing the speech intensity while ignoring the spurious background artifacts and noise. We used Symlets6 (sym6) wavelet with 6 levels of decomposition for computing the speaker energy. 
\vspace{-2mm}
\section{Measuring Dominance in a PLTL Session}
\label{sec:dominance}
\vspace{-2mm}
Dominance is a fundamental aspect of interactions in a PLTL session or small-group meeting. Authors measured the dominance in meeting using speaker diarization techniques~\cite{hung2008estimating}. A supervised model for dominance using short-utterances was developed in~\cite{basu2001learning}. However, this model was developed and evaluated on a constrained setting that was very different from the real-life scenarios such as PLTL sessions. Authors used multi-modal features derived from audio and video streams for analyzing the dominant person in a meeting segment~\cite{hung2007using}. The speaking time of speakers was found to be correlated with perceived dominance of individuals in group settings~\cite{mast2002dominance}. We developed an unsupervised feature for measuring dominance. A dominance score (DS) was assigned to each student in a PLTL session by unsupervised acoustic analysis of their speech-segments. The proposed DS encapsulates the probability of a given student to be dominant in collaborative problem-solving. We considered three features derived from speech corresponding to each speaker. This information is available from the proposed informed HMM-based diarization system as shown in Figure~\ref{fig_pltl}. The three features are turn-taken-sum ($turns$)~\cite{larrue1993organization}, speaking-time-sum ($spts$), and speaking-energy-sum ($spens$). The turn-taken-sum ($turns$) is the number of turns taken by the speaker in a given segment. A conversation turn was decided by a speech segment from the speaker cascaded between speech from other speakers and/or between speech pauses (non-speech). The speaking-time-sum ($spts$) is the sum of length of time-segments (in seconds) for which the speaker was speaking. The overlapped speech was not taken into account for estimation of speaking-time-sum ($spts$). Speaking-energy-sum ($spens$) is defined as the energy of all speech segments belonging to that speaker. The energy was computed using Wavelet Packet Decomposition (WPD)~\cite{wickerhauser1991lectures} as discussed in Section~\ref{sec:wpd}.
%
%
These features are correlated among themselves. For example, a person who is taking many turns is likely to speak for longer time than others. Also, adding the speaker energy for a longer duration will give higher $spens$. After extracting all the three features, $turns$, $spts$ and $spens$, we normalized each feature dimension. The mean and variance were calculated over the entire PLTL session (70 minute audio). We projected these normalized features onto eigen vector corresponding to the highest eigen value of the feature space. This was realized by principal component analysis (PCA) that combined the three features into a single feature, named $comb$ feature (short for "combined feature"). Let us denote the $comb$ feature by $p$. We divided the entire PLTL session into 5-minute segments. We computed the $comb$ feature for each speaker in a given segment of a PLTL session. A dominance score was estimated for each speaker in each 5-minute segment. Let us say, $p_{i}$ is the $comb$ feature corresponding to $i-th$ speaker. For the CRSS-PLTL corpus we have 6 to 9 speakers in a PLTL session including team leader. We define $comb$ feature vector $\mathbf{p} = [p_{1}, p_{2},..,p_{N}]$, where $N$ is the number of speakers. The dominance score (DS) for each speaker is estimated by passing the $comb$ feature vector, $\mathbf{p}$, through a soft-max function that converts these numbers into probability scores. Thus, we have
\begin{equation}
DS_{i}= \frac{e^{p_{i}}}{\sum_{j=1}^{N}e^{p_{j}}}, 
\label{eqnd2}
\end{equation}
for $i=1,2,..,N$; where$DS_{i}$ is the dominance score of $i-th$ speaker. For PLTL groups in particular, it is interesting to note that the dominance score of each students is an important metric with respect to inter-session variability of that group. From the previously studied supervised dominance models that predicted only the most dominant speaker, such a comparison would not be possible~\cite{jayagopi2009modeling,huang2006advances,hung2010estimating}. 
\begin{table*}[!t]
\centering
\caption{The parameters set for proposed system.}
\begin{tabular}{|c|c|}
\hline
\textbf{Parameter} & \textbf{Value}\\
\hline
Stacked DAE input layer dim.&1001\\ \hline
Stacked DAE second layer dim.&91\\ \hline
Stacked DAE bottleneck layer dim. &21\\ \hline
Number of Hidden Layers&3\\ \hline
First Layer activation&tanh\\ \hline  
Hidden Layer activation&sigmoid\\ \hline  
Initial states in HMM&12-18\\ \hline
Number of GMM components& 2-5\\ \hline
Minimum duration for HMM states& 0.5s-1s\\ \hline
Splicing context (past) & 5 frames\\
\hline
Splicing context (future) & 5 frames\\
\hline
Feature type&13-MFCC\\
\hline
Window-length&25ms\\\hline
Skip-rate&10ms\\ \hline
Sampling rate &8000 Hz\\ \hline

\end{tabular}
\label{table_params}
\end{table*}
\vspace{-2mm}
\section{Results \& Discussions}
\label{sec:results}
\vspace{-2mm}
\subsection{DER Evaluation}
\label{sec:der}
\vspace{-2mm}
We obtained the manual annotations for speech activity detection (SAD) and speaker diarization. The evaluation set consisted of one PLTL session with seven students. It was organized for approximately 70 minutes. We downsampled the audio data to 8 kHz before processing it. We used Diarization Error Rate (DER) for evaluating the proposed system. NIST Rich Transcription Evaluation~\cite{nistder} defined DER as follows: 
\begin{equation}
DER = \frac{\epsilon_{fa} + \epsilon_{miss} + \epsilon_{err} }{\epsilon_{total}},
\label{eqn_der}
\end{equation}
where $\epsilon_{fa}$ is the sum of duration of non-speech segments detected as speech, $\epsilon_{miss}$ is the total duration of speech segments detected as non-speech, $\epsilon_{err}$ is the total duration of speech that was clustered as incorrect speakers, and $\epsilon_{total}$ is the total duration of speech from the ground-truth. The parameters of the proposed system are given in Table~\ref{table_params}. We extracted 13-dimensional MFCC features from each of the seven parallel streams of a PLTL session. We chose a PLTL session with 7 students and hence 7 streams of audio data for evaluation. After concatenating the features from each stream we get a feature super-vector of dimensions 91 (=13*7). After splicing the feature super-vectors with 5 frames of past and future context as shown in Figure~\ref{fig_pltl}, we get the final dimension of features as 1001 (=11*91). The spliced feature super-vector is fed to stacked denoising autoencoder (DAE) for extracting the bottleneck features of dimension 21. A stacked DAE with three hidden layers was chosen where the middle hidden layer acts as bottleneck layer. The bottleneck (BNF) features were fed to the informed HMM-based diarization system. We used the Oracle SAD in the proposed system to validate the accuracy of the diarization system. However, we performed another case-study by formulating non-speech as an additional HMM state. We compared the diarization accuracy of BNF and raw 13-dimensional MFCC features. The concatenation of features from multi-stream was done in case of MFCC. Table~\ref{table_results} shows the diarization accuracy in various cases. The NO SAD case refers to not using any SAD labels and modeling non-speech as an additional HMM state. We knew that the non-speech has several distinct varieties, such as silences(with extreme noise), overlapped speech~\emph{etc.}. This makes it a challenging task without SAD labels. It lead to degradation in diarization accuracy (see Table~\ref{table_results}). We can see the BNF combined with HMM is robust with respect to change in minimum duration constraints and to some extent is robust to absence of SAD labels. The state-of-the-art LIUM baseline~\cite{meignier2010lium} is borrowed from our earlier work for comparison~\cite{dubey2016interspeech}. We can see an absolute improvement of approximately 27\% in terms of DER over the baseline LIUM system and approximately 12\% improvement is due to BNF features instead of using MFCC (Oracle SAD, 1s case).
%
\begin{table*}[!t]
\centering
\caption{Comparison of the DER for various parameters of the proposed system. The 13-dimensional MFCC features from each steam were concatenated for training HMM system as an additional cases for comparative study. $I_{K}$ is number of initial clusters, $t_{min}$ (s) is minimum-time HMM has to stay in each state. $I_{G}$ is the number of Gaussian used for modelling initial segments.}
\label{table_results}
\centerline{
\begin{tabular}{*{3}{|c|c|c}}
\hline
\textbf{SAD}&$feature$&$t_{min}(s)$&$I_{K}$&$I_{G}$&DER(\%)\\\hline
Oracle&21-DAE&1&12&2&\textbf{8.05}\\ \hline
Oracle&21-DAE&0.5&12&2&8.87\\ \hline
NO SAD&21-DAE&1&12&2&\textbf{15.83}\\ \hline
NO SAD&21-DAE&0.5&12&2&16.64\\ \hline
Oracle&13-MFCC&1&12&2&19.98\\\hline
Oracle&13-MFCC&0.5&12&2&18.95\\\hline
NO SAD&13-MFCC&1&12&2&33.23\\ \hline
NO SAD&13-MFCC&0.5&12&2&41.71\\ \hline
LIUM~\cite{meignier2010lium} & &  & & &35.80 \\ \hline
\end{tabular}
}
\end{table*}
%
\vspace{-2mm}
\subsection{Dominance Score}
\label{sec:4p2}
\vspace{-2mm}
The PLTL session was divided into segments of five minutes' duration. We choose a PLTL session with seven students. For each five-minute segment, we compute a dominance score ($DS$) for each of the seven students using unsupervised acoustic analysis. We conducted Intelligent Listening Test (ILT) for annotating each five-minute segment of the PLTL session by assigning a ground-truth dominance rating ($D_{rate}$) for each student per segment. Three annotators listened to each five-minute segment and assigned a dominance rating ($D_{rate}$) for each student per segment. The ground-truth dominance rating, $D_{rate}$, was a number between 1 and 5. The speakers who were present in the whole session but did not speak in the chosen segment were assigned a dominance rating, $D_{rate}=1$. The scores of $D_{rate}=2$ and $D_{rate}=5$ were assigned to the least-and most-dominant students who spoked in that segment. For students who spoke in that segment and were neither least-dominant nor most-dominant, we assigned them a $D_{rate}$ between 2.25 and 4.75. It was possible to score 2.25, 2.50, 2.75, 3.0, 3.25, 3.50, 3.75, 4.0, 4.25, 4.50 and 4.75. However, no fractions other than these were used to ensure consistency in evaluations. We averaged the ground-truth rating ($D_{rate}$) of all three listeners to get a final ground-truth that was used for computing the correlation with unsupervised dominance score ($DS$). Since the proposed dominance score, $DS$ was derived using unsupervised acoustic analysis, we used Pearson's correlation between ground-truth dominance rating ($D_{rate}$) and proposed dominance score ($D_{rate}$). The correlation between ground-truth $D_{rate}$ and proposed $DS$ was 0.8748. The high correlation value validates the efficacy of proposed dominance score, $DS$, for characterizing individuals in a PLTL group. 
\vspace{-2mm}
\section{Conclusions}
\vspace{-2mm}
This paper showed that the acoustic cues could help in mining meaningful analytics such as dominance from a Peer-led Team Learning (PLTL) session. We used stacked denoising autoencoder (DAE) for dimension reduction of spliced feature super-vectors obtained by concatenating features from seven streams of multi-stream PLTL data. The bottleneck (BNF) features from stacked DAE were fed to an informed HMM-based speaker diarization system. Finally, the dominance score was estimated using unsupervised acoustic analysis of each speaker segment. We evaluated the proposed system on CRSS-PLTL Corpus established in~\cite{dubey2016interspeech}. 
\bibliographystyle{IEEEbib}
\bibliography{strings,refs}
\end{document}